\documentclass[aps,prc,twocolumn,showkeys,showpacs]{revtex4-1}
\usepackage{amssymb}
\usepackage{graphicx}
\usepackage{color}
\begin{document}
\title
   {\bf Fission dynamics of $^{252}$Cf}
\author{A. Zdeb, A. Dobrowolski, M. Warda}

\affiliation{Department of Theoretical Physics, Maria Curie-Sk\l odowska University, Lublin, Poland}
\date{\today}

\begin{abstract}
The time-dependent generator coordinate method with the gaussian overlap approximation (TDGCM+GOA) formalism is applied to describe the fission of $^{252}$Cf. We perform analysis of fission from the initial states laying in the energetic range from the ground state to the state located 4 MeV above the fission barrier. The fission fragment mass distributions, obtained for different parity, energy of levels and types of mixed states, are calculated and compared with experimental data. The impact of the total time of wave packet propagation on the final results is studied as well. The weak dependence of obtained mass yields on the initial conditions is shown.
\end{abstract}

\maketitle

\section{Introduction}

The properties of fission  and  its residues play a significant  role for  the  nuclear  power  industry  and  the  related  fields (e.g. nuclear waste management). One of the main observables of fission - fragment mass distribution - is an important input in the r-process calculations which allow to explain the abundances of isotopes in the Universe. An extensive experimental and theoretical studies  of this phenomenon (see e.g. reviews~\cite{CW,KP,R,VH73}) are carried on since the first evidence of fission has been observed~\cite{FP}.\\
Proper description of fission fragment mass distribution is still a challenging task for nuclear theory. Several models have been developed so far to describe experimental observations. The first theoretical explanation of fission was given on the ground of the liquid drop model. The  fission fragment mass distributions calculated within this method are symmetric, as a consequence of ignoring microscopic effects \cite{NS65}.\\
A more sophisticated approach - statistical model \cite{W76} - allows to determine the probability of division of nucleons between nascent fragments at the scission point. The calculated mass distributions are overestimated in comparison with the experimental ones.\\
The distribution of fission fragments and the mean value of total kinetic energy (TKE) can be obtained using the the improved scission point model~\cite{A13}. This is an extension of the previous approach~\cite{W76} and it allows to calculate interaction energy between fragments and deformation energy of the scission point configuration.\\
The microscopic scission point method~\cite{PAN12} is based on the analysis of deformation and mass asymmetry of fragments that may be created after scission. The energies are computed within microscopic self-consistent model. The assumed deformations and masses allow to calculate the total kinetic energy of fragments and the probability that the certain mass asymmetry would be observed.\\ 
Fission fragment mass distributions may be also obtained using Langevin formalism~\cite{pr1,pr2,AC13}. This approach allows to include plenty important dissipation and pairing effects in description of fission process. Additionally, it is also possible to estimate the fission time scale in this model. Such calculations are performed under the assumption of the same deformation of both fragments.\\
Interesting results were obtained within the similar approach that treats the nuclear shape evolution as Brownian motions of nucleons. The possible directions in the multi-dimensional deformation space and their statistical weights are found using Metropolis method \cite{JR1, JR2}. Although the model reproduces experimental data with high accuracy, it contains phenomenological parameter - {\it critical radius constant} which value results much on the accuracy in data reproduction.\\
The authors of the general description of fission observables (GEF method)~\cite{KHS,kod} obtained very good agreement with observed mass yields of most of measured fissioning isotopes. In this model the macroscopic potential energy surface (PES) is corrected by adding shell effects which are simulated by parabolic potential.\\ 
There were also several attempts to describe fission fragment mass distributions in a fully  microscopic way, i.e. using  the Hartree-Fock-Bogolubov (HFB) method with the Skyrme energy functional \cite{Schunck} or the time-dependent Hartree-Fock (TDHF) method~\cite{HG2005, D8}, explained more detailed in the next section. The pure self-consistent models (HFB) produce only the most probable fragment mass asymmetry~\cite{WSN}. Dynamic effects, added on the top of the static results, are essential to obtain the full fragment mass distribution.\\
Recently the time-dependent density functional theory (TDDFT) was applied to calculate TKE and mass yields of $^{258}$Fm~\cite{lx}. The authors analyzed fluctuations in
scission time, pre- and post-scission emissions of neutrons and protons. Also the correlations between TKE and collective deformation of daughter nuclei were studied. The obtained mass yield is too narrow in comparison to the measured one.\\
 The aim of the present work is to examine how the initial conditions affect the obtained mass yield, especially how this quantity depends on the  excitation energy and the parity of the initial state. We studied fission of $^{252}$Cf using the TDGCM+GOA approximation explained in detail in Refs.~\cite{HG2005,D8,RDV16,BGG} and briefly described in the next Section. In Section II we present the theoretical framework. Section III contains results of our investigations. 
Conclusions are presented in the last section.

\section{Method}

\begin{figure}[h!]
 \includegraphics[height=0.93\columnwidth,angle=270]{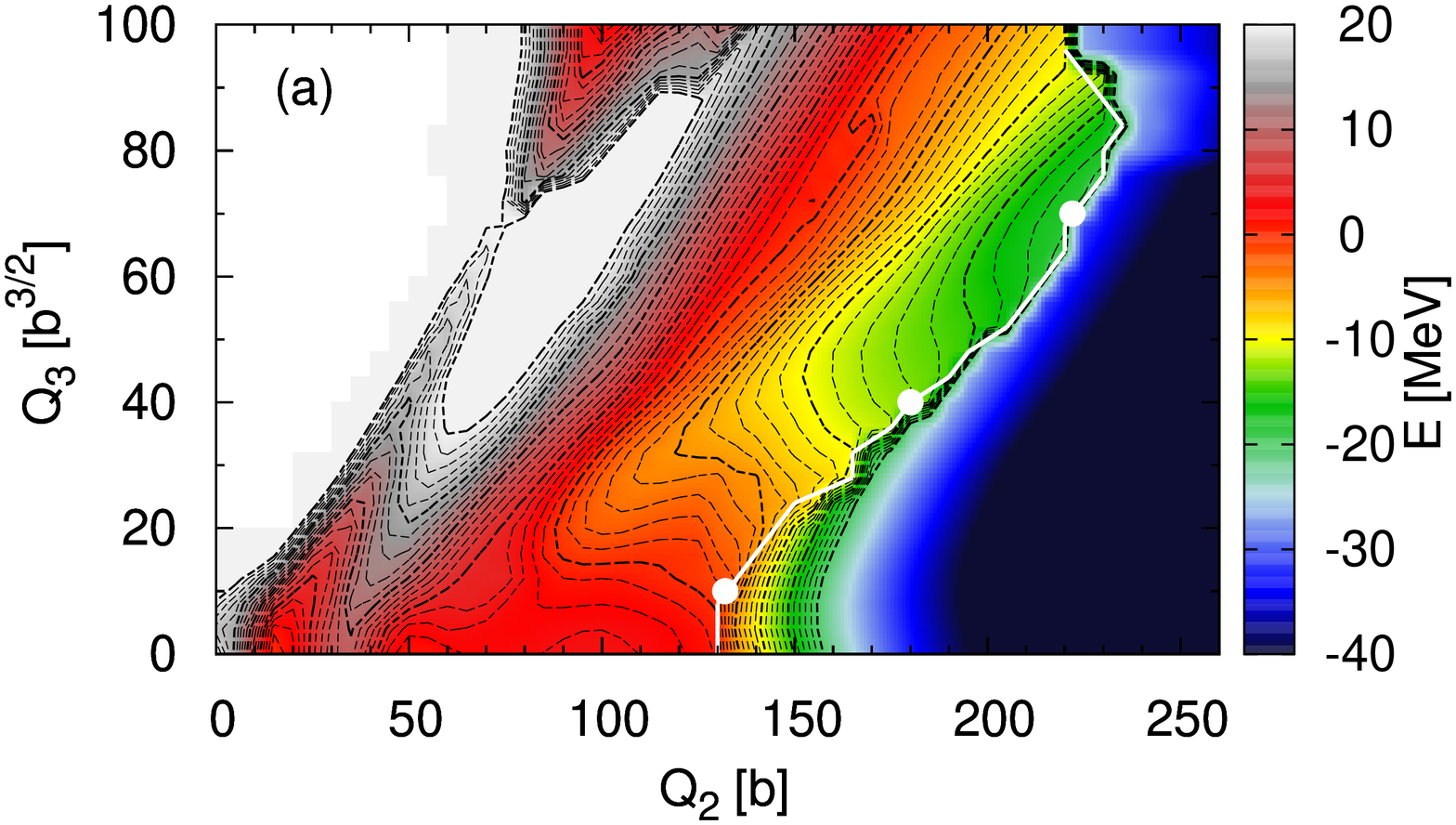}\\
 \includegraphics[height=0.55\columnwidth,angle=90]{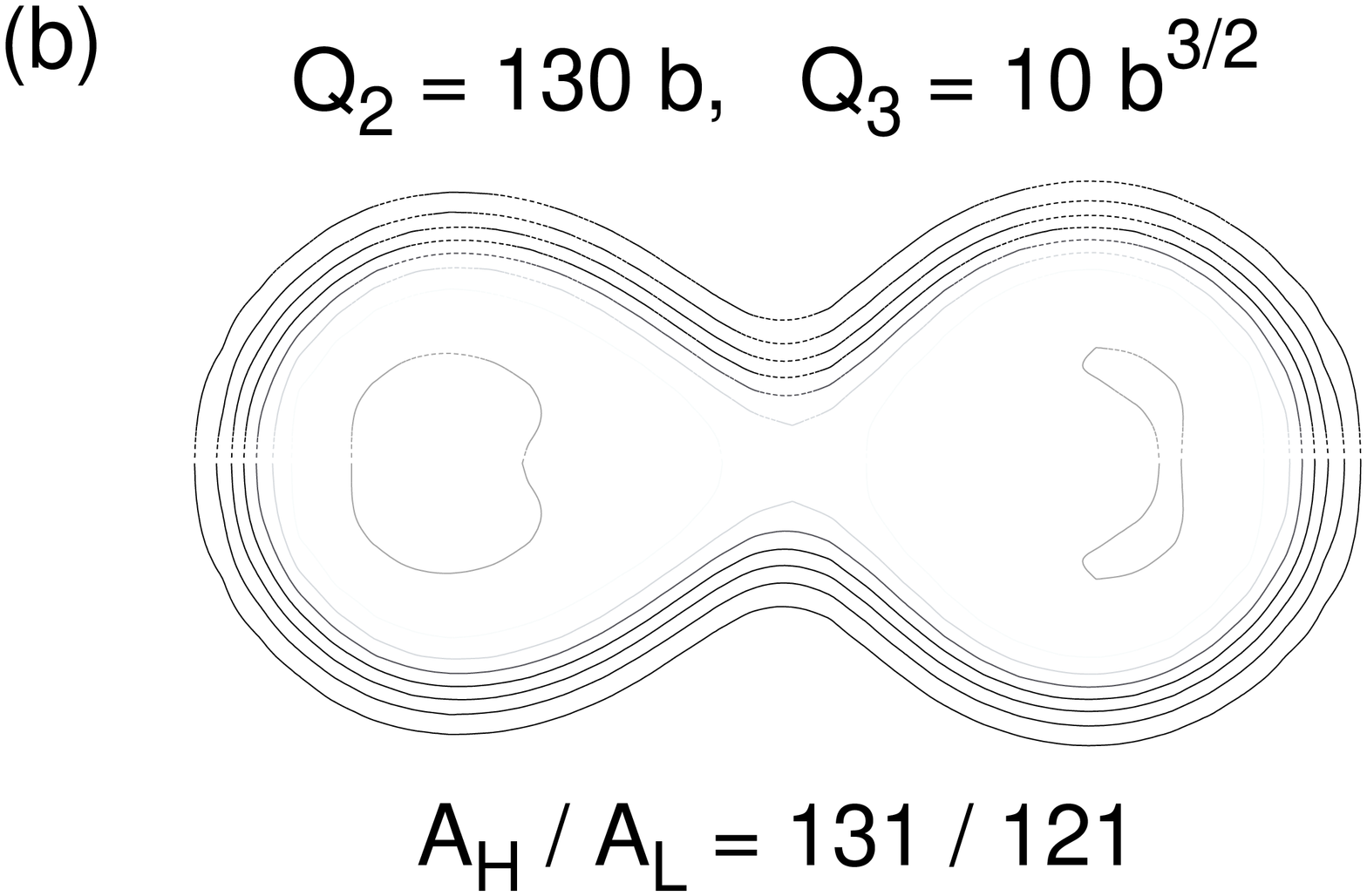}\\
 \includegraphics[height=0.55\columnwidth,angle=90]{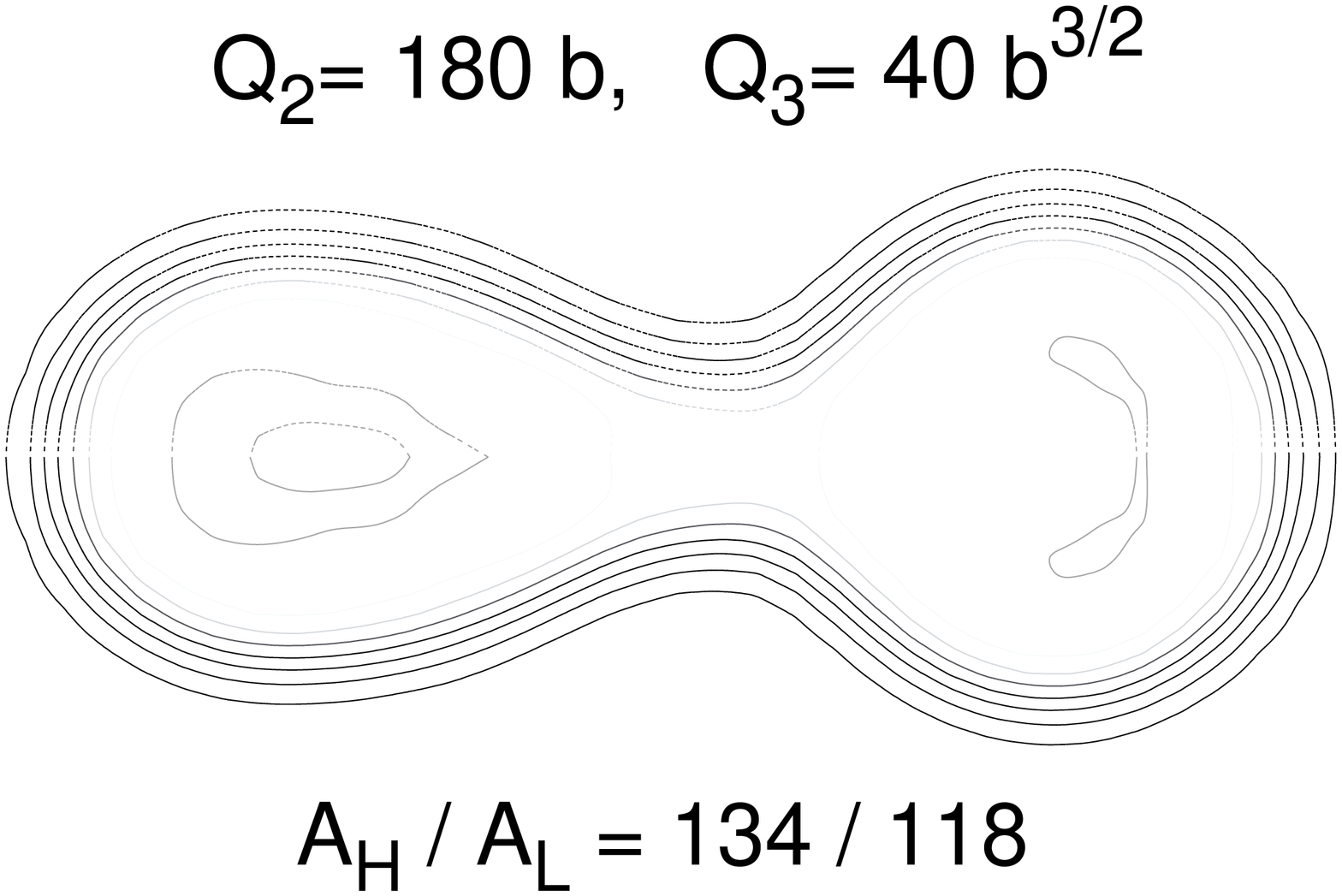}\\
 \includegraphics[height=0.55\columnwidth,angle=90]{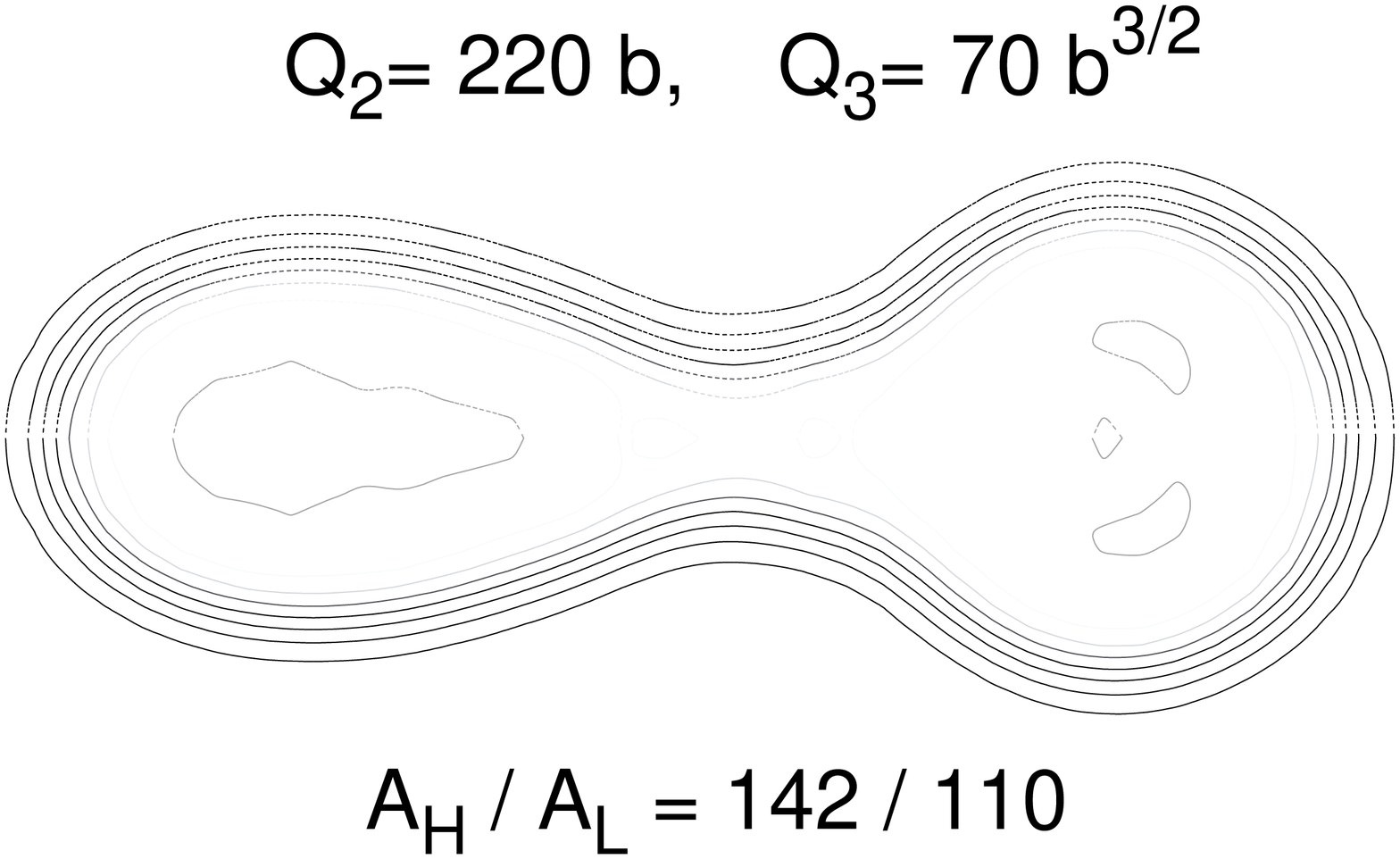}\\
\caption{(a) The potential energy surface of $^{252}$Cf. The p-sl is depicted in white. (b) Density profiles corresponding to the configurations marked by dots are presented in panel.}
\label{pes}
\end{figure}

We perform our analysis in two steps: (I) static calculations of the PES and mass parameters, and (II) dynamic part of the evolution of the wave packets. (I) The PES is calculated using the Hartree-Fock-Bogolubov model with the D1S Gogny-type interactions. The HFB equations are solved with constraints on quadrupole $Q_{20}$ and octupole $Q_{30}$ moments of the total nuclear density. Details of the calculation can be found in Refs.~\cite{WE1,WR,WE}. The mass parameters are computed within the adiabatic time-dependent Hartree-Fock (ATDHF) formalism~\cite{RENP,RG}. We analyze the evolution of the wave packet from the initial state up to the rupture of a nucleus into two fragments. Therefore we need to determine possible scission configurations with two touching daughter nuclei. In practice, self-consistent calculations allow to find the pre-scisson shape of a nucleus as the last point in the deformation space before the rupture of the neck~\cite{D12}. Most of the fission fragments' properties are determined between saddle configuration and this point. The set of these points referring to various octupole deformations creates pre-scission line (p-sl) which is presented in Fig.~\ref{pes}a with the white solid line. 
(II) The dynamic calculations are done within the TDGCM+GOA formalism. The theoretical framework of this method is explained in detail in Refs.~\cite{HG2005,D8,RDV16} and references therein.\\
The main constituent of the model is the collective Hamiltonian which is taken in the form: 

\begin{equation}
\begin{array}{r}
\widehat{H}_{\rm coll}=-\frac{\hbar^2}{2\sqrt{\gamma}}\sum\limits_{i,j=2}^3\frac{\partial}{\partial_{Q_{i0}}}\sqrt{\gamma}B_{ij}(Q_{20},Q_{30})\frac{\partial}{\partial_{Q_{j0}}}\\
+V(Q_{20},Q_{30}),
\end{array}
\label{Hcol}
\end{equation}
where $B_{ij}$=$\mathcal{M}^{-1}$ are the mass parameters of the collective inertia tensor $B$ with $\mathcal{M}^{-1}$ defined as:
\begin{equation}
\begin{array}{r}
\mathcal{M}_{i,j}=\sum\limits_{i,j=2,3}=(M^{(-1)}_{ik})^{-1}(M^{(-3)}_{kl})^{-1}(M^{(-1)}_{lj})^{-1}.
\end{array}
\label{m}
\end{equation}
The moments of order $-m$ are given by:
\begin{equation}
\begin{array}{r}
M_{i,j}^{(-m)}=\sum\limits_{\mu \nu}\frac{\left < \Phi(Q_{20},Q_{30})|\widehat{Q}_{i0}|\mu \nu\right >\left <\mu \nu|\widehat{Q}_{j0}|\Phi(Q_{20},Q_{30})\right >}{(E_{\mu}-E_{\nu})^m},
\end{array}
\label{mm}
\end{equation}
where $\Phi(Q_{20},Q_{30})$ are solutions of the constrained HFB variational principle and $\left |\mu \nu \right >$ are the quasi-particle states with energies $E_{\mu}$ and $E_{\nu}$. The quantity $\gamma$ is the determinant of a metric tensor in the two-dimensional space of collective variables $(Q_{20},Q_{30})$. 

Recently the improved approach to the moments of inertia calculations has been used in the fission barrier penetration analysis~\cite{bn}. The non-perturbative mass parameters were obtained in the $Q_{20}-Q_{22}$ deformation space. It has been shown that minimization of action integral with the non-perturbative mass parameters modifies the fission trajectory in the barrier region. The penetration probability is higher in comparison to that resulting from dynamic calculations within perturbative inertias. This is an important constituent of the fission half-lives studies. However, the distribution of the probability current along the p-sl depends essentially on the evolution directions beyond the barrier, rather than the trajectory of the system through the saddle~\cite{n}. The values of perturbative and non-perturbative masses differ around the ground state minimum but at large elongations both approaches produce fairly similar collective parameters~\cite{bn22}. Thus fission mass yields should not be strongly modified when the perturbative "cranking" inertias were replaced by the non-perturbative ones.

To find the initial collective wave function of the $n-$th state $g^{\pi}_{\rm n}(Q_{20},Q_{30},t=0)$ with parity $\pi$, the eigenproblem $\widehat{H}_{\rm coll}g_{\rm n}^{\pi}=E_{\rm n}g_{\rm n}^{\pi}$  is solved in the two-dimensional  ground state well ($0\le Q_{20}\le 55$ b, $-40\le Q_{30}\le 40\,\rm b^{3/2}$). Since eigenstates of the Hamiltonian of the mother nucleus are stationary, the initial collective wave functions (for $t=0$) are generated in the ground state well $V'(Q_{20},Q_{30})$ that is slightly modified HFB potential $V(Q_{20},Q_{30})$. The bottom of the potential well stays unchanged while the region beyond the barrier of $V'(Q_{20},Q_{30})$ is generated by linear extrapolation to large values of energy. The wave packets obtained in this way may be treated as the eigenstates of fissioning nucleus and the procedure of time evolution through the realistic nuclear potential may be efficiently applied. \\
The dynamic part of calculations is based on the numerical code originally developed by Goutte {\it et al.}~\cite{HG2005} and later on enhanced by the implementation of determinant of the metric tensor $\gamma$. The probability flux $\vec J(Q_{20},Q_{30},t)$ flowing through each point along the p-sl coordinates ($Q^{\rm sc}_{20},Q^{\rm sc}_{30}$) (see Fig.~\ref{pes}) is defined as follows:

\begin{equation}
\begin{array}{r}
\vec J(Q_{20},Q_{30},t)=\frac{\hbar}{2\dot{\imath}}\sqrt{\gamma}B(Q_{20},Q_{30})\times\\[+0.5ex]
\lbrack g^*(Q_{20},Q_{30},t)\nabla g(Q_{20},Q_{30},t)-\\
g(Q_{20},Q_{30},t)\nabla g^*(Q_{20},Q_{30},t)\rbrack.\\
\end{array}
\label{j}
\end{equation}
One can therefore obtain the probability that the fissioning system reaches a certain point in a deformation space by the expression:
\begin{equation}
\begin{array}{l}
P(Q^{\rm sc}_{20},Q^{\rm sc}_{30})=\int\limits_{t=0}^{t=T^{\rm propag}} \vec J(Q^{\rm sc}_{20},Q^{\rm sc}_{30},t)\cdot \vec n \,\ dt.
\end{array}
\label{jc}
\end{equation}
Above, $\vec{n}$ stands for the normal to the p-sl vector in the point $(Q^{\rm sc}_{20},Q^{\rm sc}_{30}).$ In this formula $T^{\rm propag}$ is a time of propagation which will be discussed in Section III~A.\\
Each point of the p-sl corresponds to a different molecular shape of a nucleus. Few illustrative configurations are shown in Fig.~\ref{pes}b. As it was mentioned above, from the dynamic part of the calculations one can get the probability $P(Q^{\rm sc}_{20},Q^{\rm sc}_{30})$ that a nucleus takes a certain shape before splitting. It is usually assumed that the neck rupture takes place at $z$ coordinate where the neck is the thinnest. Nevertheless, this is a rather simplified picture ignoring all possible fluctuations caused by collective nuclear surface vibrations. These effects may be included e.g. by the gaussian smoothing~\cite{RDV16} or by applying the random neck rupture  (rnr) mechanism~\cite{B83,B88,B90,WZ15}. In the latter method, for each pre-scission shape corresponding to the scission configuration $(Q_{20}^{\rm sc},Q_{30}^{\rm sc})$ the probability of splitting of a nucleus at the certain position on the symmetry axis $OZ$ along the neck is evaluated. Thus the mass distribution is given by:

\begin{equation}
P(Q_{20}^{\rm sc},Q_{30}^{\rm sc})=\exp[-2\gamma\sigma(z)/T].
\label{Pfrag}
\end{equation}\\
Here $\sigma(z)= 2 \pi \int_0^{\infty}  r_\perp \rho(z, r_\perp) dr_\perp$ is a linear density of a nucleus along the symmetry axis $z$, $T$ is a temperature of a nucleus in at pre-scission deformation and $\gamma$ is the surface tension coefficient with a standard parametrization given in Ref.~\cite{BS}. The temperature $T$ is of Boltzmann form and depends on excitation energy $E^*$: $T=\sqrt{12E^{*}/A}$. The energy $E^*$ is defined as a difference between the eigenenergy $E_{\rm n}$ of a propagated state $g_{\rm n}^{\pi}$ and the potential energy of a nucleus at the pre-scission point: $E^*=E_{\rm n}-E^{\rm sc}_{\rm HFB}$. This is a standard parametrization of the rnr model without any fitting procedures. However a possible modification of the $\gamma/T$ ratio affects the broadness of the mass yields~\cite {WZ15}. The final fragment mass distribution is obtained as a convolution of the density current probability distribution along the p-sl and the rnr mechanism.

\section{Results}

\begin{figure}[h!]
 \includegraphics[height=1.5\columnwidth,angle=0]{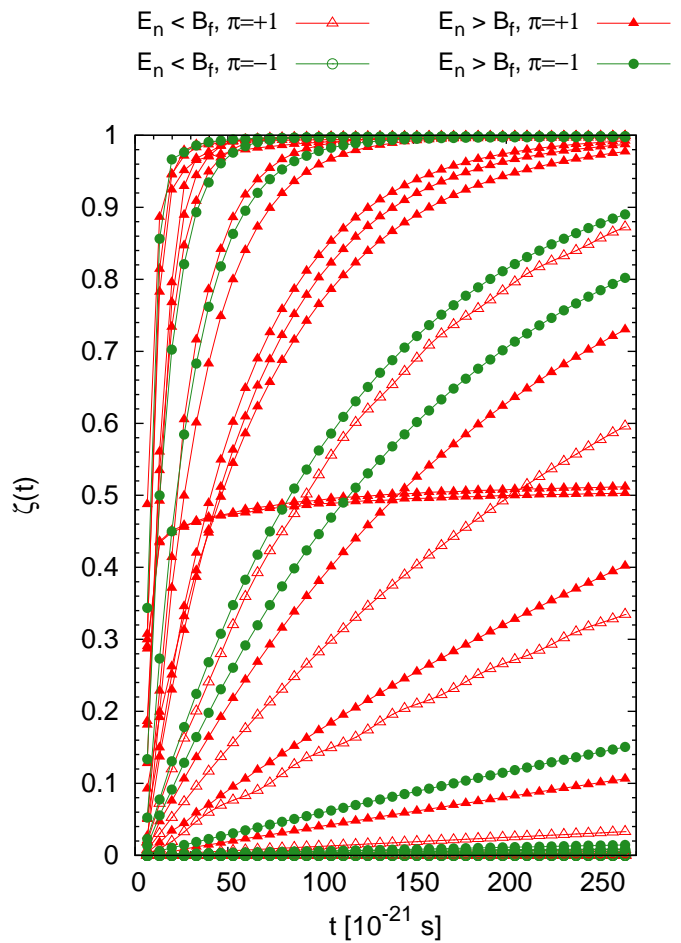}
\caption{The tunneling probability $\zeta(t)$ (Eq.~\ref{dzeta}) of $^{252}$Cf eigenstates ($g^{+}$ - red triangles, $g^-$ - green circles) as a function of time. The open symbols refer to the states located under fission barrier while the full ones correspond to the states laying above the barrier.}
\label{fig1}
\end{figure}

We have chosen the neutron-rich $^{252}$Cf isotope for our investigations which represents an asymmetric type of fission. This nucleus has been extensively studied experimentally and many important observables such as mass, TKE distribution and average prompt neutrons multiplicities are measured~\cite{CF,c1,c2,c3,CFn} and available for comparisons with theoretical predictions.  The fission barrier of $^{252}$Cf, calculated within the HFB model with Gogny forces D1S, is equal to $B_{\rm f}$=9.71~MeV. In the present work we investigate the eigenstates of the collective Hamiltonian~(\ref{Hcol}) with the  potential $V'(Q_{20},Q_{30})$ located in the energetic range from the ground state to $B_{\rm f}$+4 MeV.

\subsection{Propagation time}
To avoid the reflections at the edges of the $(Q_{20},Q_{30})$ grid we apply the absorbing complex potential~\cite{vim} which is active in the region beyond the scission line. Since the wave packet is absorbed after crossing the p-sl, it is possible to calculate the reduction of the density probability $\zeta(t)$ in the considered collective space $(Q_{20},Q_{30})$ at each time step:
\begin{equation}
\zeta(t)=1-\int|g^{\pi}(q_{20},q_{30},t)|^2dq_{20}dq_{30}.
\label{dzeta}
\end{equation}
\begin{figure}[h!]
 \includegraphics[height=1.05\columnwidth,angle=270]{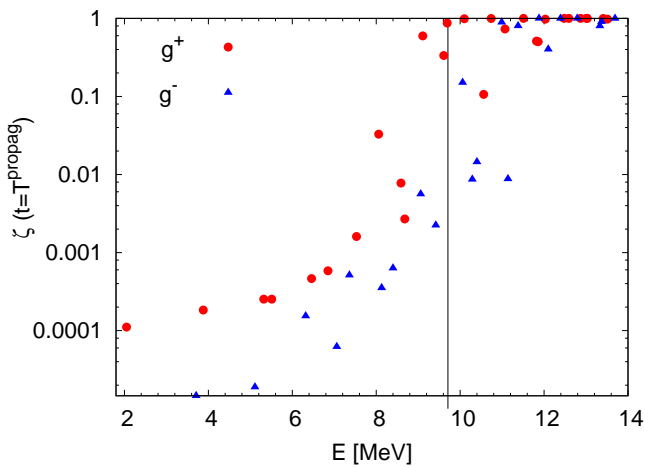}
\caption{The energy dependence of the tunneling probability $\zeta(t=~T^{\rm propag})$ (Eq.~\ref{dzeta}) of $^{252}$Cf eigenstates $g^{\pi}$. The fission barrier height $B_{\rm f}$=9.41 MeV is marked by the vertical black line.}
\label{fig2}
\end{figure}
\begin{figure}[h!]
 \includegraphics[height=1.05\columnwidth,angle=270]{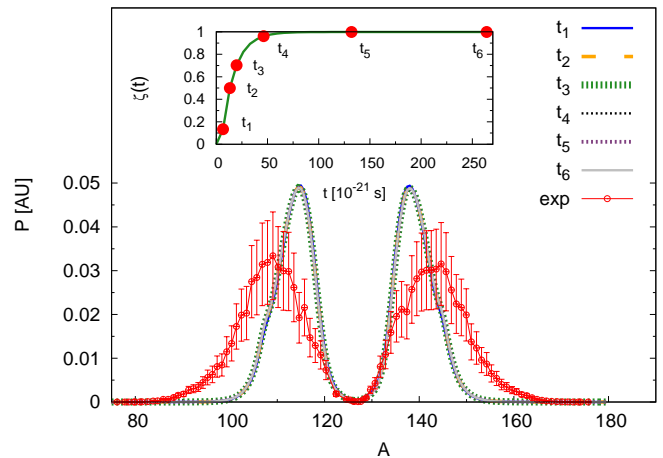}\\
\caption{The tunneling probability $\zeta(t)$ (Eq.~\ref{dzeta}) of eigenstate $n=~36, \pi=-1$ as a function of time (inner panel) and the fission fragment mass distributions calculated for denoted times of propagation $(t_1,...,t_6)$. The experimental data were taken from \cite{CF} and corrected by an average number of emitted prompt neutrons \cite{CFn}.}
\label{fig3}
\end{figure}
This quantity gives the information about the survival rate against fission after time $t$. It may be also interpreted as a tunneling probability of a particular state through the fission barrier.\\
 In Fig.~\ref{fig1} changes of $\zeta(t)$ as a function of time for each of the considered eigenstates of a mother nucleus with $E_{\rm g.s.} \le E \le B_{\rm f}$+4 MeV are displayed while Fig.~\ref{fig2} shows the dependence of the value of $\zeta(t=T^{\rm propag}$) on the energy of an initial state. There is a visible tendency that the states with higher energy propagate faster - rapid increase of $\zeta(t)$ may be observed. The lowest states propagate very slowly and after $T^{\rm propag}$ fission probability is negligibly small - below 1\%. In the same time levels with $E_{\rm n} > B_{\rm f} + 2$ MeV propagate rapidly and at $t=50\cdot 10^{-21}$ s and $\zeta(t)$ saturate at value close to 1. It means that the wave function completely run away from the ground state well. Several levels laying around $E_{\rm n} \approx B_{\rm f} \pm 2$ MeV have a small tunneling probability - after $T^{\rm propag}$ only about 1\% of the wave packet crossed p-sl. The other levels from this range propagate much faster and after $T^{\rm propag}$leave the vicinity of the ground state. There are 2 eigenstates ($g_{34}^+$, $E_{34}=11.82$ MeV; $g_{35}^+$, $E_{35}=11.86$ MeV) which behavior diverges from this main tendency - after fast reduction of the density probability at the very beginning of the time evolution they stabilize at some value of $\zeta(t)$. The fission probability $\zeta(t)$ changes its value just by around $0.03 - 0.05$ when the time of propagation is extended twice.\\
 Since such a different behavior of individual states is observed, we decided to check whether the termination of time evolution at a certain moment influences much on the final mass distribution. Fig.~\ref{fig3} shows the mass  yields obtained for different times of propagation of the initial state $g^-_{36}$ (inset of Fig.~\ref{fig3}). The shape of mass distribution does not depend significantly on the duration of the time evolution - the curves resulting from all time steps overlap. Similar behavior is typical for all tested levels. This allows us to stop the time evolution after arbitrary chosen $T^{\rm propag}=2.6\cdot 10^{-23}$ s. Further calculations show that even if time of propagation is extended twice, the tunneling probability stays almost constant and the final distribution of the probability current density along the p-sl does not significantly change.\\
One can also observe that $\zeta(t=T^{\rm propag})$ is smaller for states with negative parity than for positive ones with similar energies. This  tendency may be understood when one looks at the shape of the PES in the ground state well and fission barrier region. The saddle is located at $Q_{30}$=0 and energy grows with increasing octupole deformation. Since the states with negative parity prefer $Q_{30}\neq 0$ channel their propagation is hindered by the potential around the saddle point.\\

\subsection{Parity dependence}

\begin{figure}[h!]
 \includegraphics[height=1.05\columnwidth,angle=270]{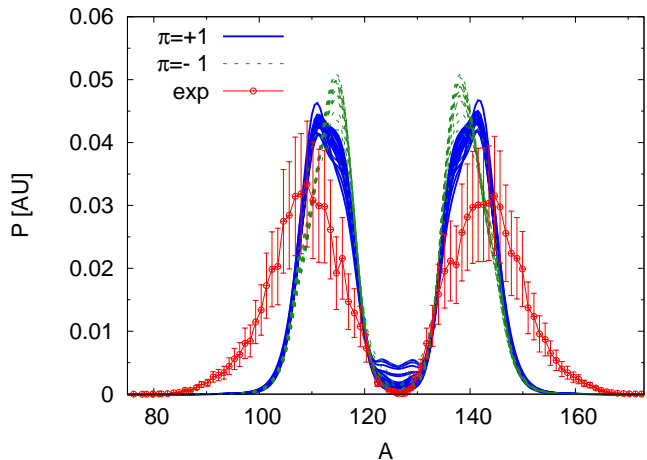}
\caption{The fission fragment mass distributions as a result of propagation of positive (blue solid line) and negative (green dashed lines) states $g^{\pi}$ in the considered range of eigenvalues.}
\label{fig4}
\end{figure}

In Fig.~\ref{fig4} we show the mass yields, as a result of time evolution of positive and negative parity states in considered energy regime, are displayed. The states of the same parity lead to very similar shapes of yields - heavy and light fragments peaks keep almost the same position independently on the initial energy. Furthermore, mass distributions have very similar broadness. The most probable masses of fragments obtained from negative parity are shifted by 2 mass units to the center of distribution in comparison to the ones resulted from the positive parity functions. Some of considered states (with positive parity and $E<B_{\rm f}$) lead to a slightly different mass distribution around $A=126$ than the others. The obtained mass yields show that in these cases the symmetric fission channel has a small contribution.

\subsection{Energy dependence}

\begin{figure}[h!]
 \includegraphics[height=1.05\columnwidth,angle=270]{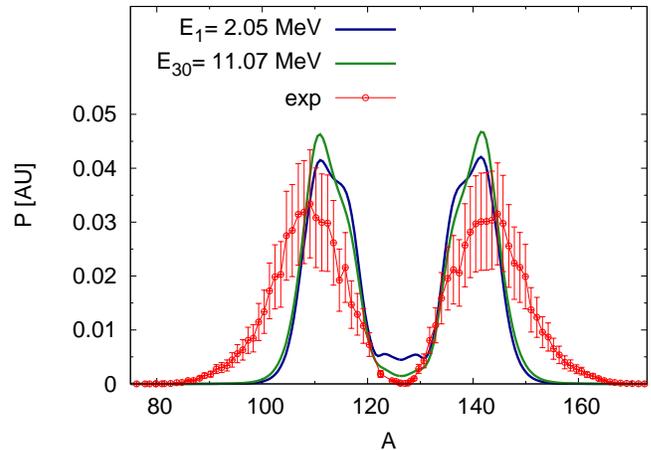}
\caption{The comparison of mass yields obtained from time evolution of the ground state at energy $E_1= 2.05$ MeV (blue line) and the state with eigenvalue $E_{30}=11.07$ MeV (green line). Both states have positive parity.}
\label{fig5}
\end{figure}

To investigate the impact of the energy of the initial state on the final fragment mass distribution we have compared the results obtained after time evolution of the lowest state $E_1=2.05$ MeV and the state laying at energy $E_{30}=11.07\,\rm MeV$ (see  Fig.~\ref{fig5}). Obtained yields have very similar shape. Self-consistent calculations show that the PES depends weakly on the excitation energy~\cite{NSW}, thus any qualitatively important changes in the fission yields should not expected. Since both initial states have the same (positive) parity, the positions of $A_H$ and $A_L$ peaks cover. In the case of $E_1$ state the symmetric-fission mode contribution is non-zero, what is not observed in the experiment. Keeping in mind that for the first eigenstate $\zeta(T^{\rm propag})<<1\%$ and the energy distance between these eigenstates is rather large, the differences between theoretical and experimental yields may be treated as negligibly small. The most probable $A_H/A_L$ ratio 140/112 is slightly smaller than the measured  - 142/110 and the calculated mass yields are narrow in comparison to the experimental one. Similar results are obtained for any initial state. We may conclude that the fragment mass distribution of a nucleus excited to energy below or around fission barrier height stays almost unchanged.

\subsection{Mixed states}
According to the well established picture of induced fission, the excited nucleus does not residue in its pure eigenstate but fissions from the state which is the superposition of its eigenstates. Measured mass yields of low-energy induced-fission does not differ significantly from spontaneous-fission ones. Thus problem of state mixing in the theoretical description of the process should be also considered. We examined various types of mixing states, using the distributions of gaussian-shape:
\begin{equation}
P(E_{\rm n})=\frac{1}{\sigma \sqrt{2\pi}}\exp\left[-\frac{(E_{\rm n}-\mu)^2}{2\sigma^2}\right]
\end{equation}
and of the Fermi-shape:
\begin{equation}
P(E_{\rm n})=\frac{1}{B_{\rm f}}\frac{1}{\left[\exp\left(\frac{E_{\rm n}-B_{\rm f}}{d}\right)+1\right]}
\end{equation}
\begin{figure}[h!]
\hskip5pt
 \includegraphics[height=1.05\columnwidth,angle=270]{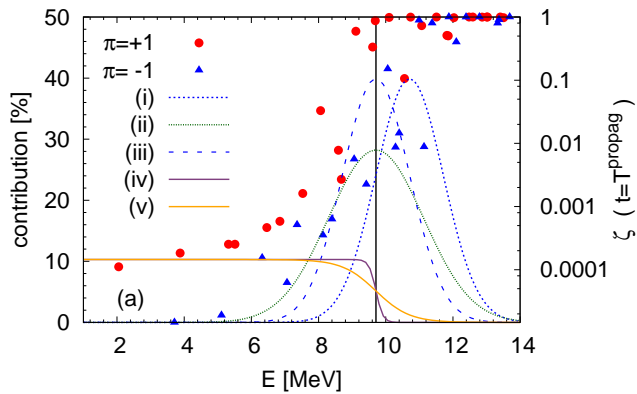}\\
\vskip20pt
 \includegraphics[height=1.05\columnwidth,angle=270]{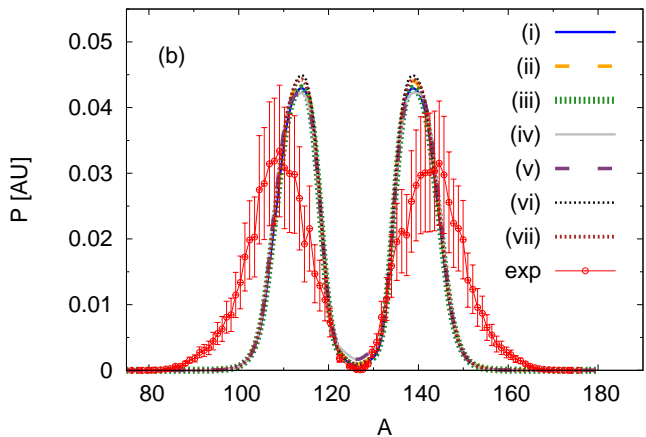}\\

\caption{Upper  panel (a): Types of statistical mixing of eigenstates and fission fragment mass distributions obtained as a results of time evolution of these wave packets (lower panel (b)). The left axis of upper panel shows the percentage contribution of each state according to particular statistical distribution used in the mixing procedure. The right axis refers to the tunneling probability $\zeta(t)$ of considered states.}
\label{fig6}
\end{figure}

Mixing of eigenstates in several energy regimes are shown in Fig.~\ref{fig6}(a).  The initial wave packets are generated as the linear combinations of all single states located in this energy regime with statistical weights given by the considered probability distributions (i)-(v). The Gaussian-type mixing are performed for (i) $\, \sigma^2=~1\, (\rm MeV)^2,\,$ $\mu=~B_{\rm f}+1 \,\rm MeV$, (ii) $\, \sigma^2=~2\,(\rm MeV)^2$, $\mu=~B_{\rm f}+2,\, \,\rm MeV,\,$ (iii) $\, \sigma^2=~1\,(\rm MeV)^2$, $\mu=~B_{\rm f}+1\, \rm MeV$ and Fermi-type with the center of distribution located at $B_{\rm f}$ and diffuseness parameters (iv) $\, d=0.1 \,\rm MeV$ and (v) $\, d=0.5\,\rm MeV$. Additionally we construct two wave functions [not shown in Fig.~\ref{fig6}(a)] as combinations with equal contributions of the eigenstates from the eigenvalues ranging between (vi) $\, B_{\rm f}\le E_{\rm n} \le B_{\rm f}+1\,\rm MeV$ and (vii) $\, B_{\rm f}\le E_{\rm n} \le B_{\rm f}+2\,\rm MeV$. In Fig.~\ref{fig6}(b) the fission fragment mass distributions obtained as the result of time evolution of the mixed wave packets are displayed and compared to the measured ones. The shapes of these mass yields show that there are no important qualitative differences between final distributions resulted from the considered statistical methods of mixing. The resulted peaks are shifted by 4 mass units in comparison to the experimental ones. One can observe that the total widths of $A_H,A_L$ peaks are now slightly broader than those resulting from the evolution of the single states. Moreover, the aforementioned symmetric mode of the low laying states does not affect the yield in Fig.~\ref{fig6}b. The influence of this effect is washed out by the dominant contribution coming from levels with higher energy and faster barrier penetration.\\

\subsection{Discussion of the results}
The present method allows to obtain the main characteristics of observed mass yield of $^{252}$Cf. The calculated most probable fragment mass asymmetry fairly reproduces experimental data. We predict also diminishing yield for symmetric mass split. 

The largest discrepancy is obtained in the most asymmetric part of the fragment mass distribution, where our results underestimate experimental evidence. This is because nuclear configurations corresponding to such asymmetry appear on the p-sl only for large values of octupole deformation: $Q_{30}^{\rm sc} > 80\, \rm b^{3/2}$. Since they lay high in energy (see Fig.~\ref{pes}a) the probability that the probability flux flowing through this region is small.

In order to obtain a better description of such asymmetric fragmentations, some extensions of the model are required. There are several options of possible modifications. One of them is an extension of the deformation space by another degree of freedom (hexadecapole moment, strength of pairing correlations), including quantal effects on a neck rupture mechanism or energy dissipation effects between the intrinsic and collective degrees of freedom. 

Studies of the three-dimensional PES of $^{252}$Cf shown that the hexadecapole moment plays an important role in the description of fission mass asymmetry~\cite{w3}. Namely, the scission configuration may be reached for lower quadrupole moment within different mass asymmetry when the PES is spanned on $Q_{20}-Q_{30}-Q_{40}$ space. Moreover, there were found several fission paths in such three-dimensional deformation space which lead to different pre-scission shapes. 

 As it was shown in the dynamical description of nuclear fission, proton and neutron pairing correlations are important ingredients~\cite{bs14}. Pairing correlations have its impact on action integral minimization. The interplay between the potential energy and collective inertia affects the propagation direction chosen by the system in the deformation space.

The detailed, microscopic analysis of the pre-scission configuration are also needed. One may investigate the single particle energies and  density distributions just before splitting~\cite{w4}. It allows to study the formation of the nascent fragments and predefine the most possible mass asymmetry. It is also possible to identify fission fragments through the localization of the nucleons wave functions~\cite{yg}.\\
 The relevance of the fluctuations in the fission dynamics also has been recently demonstrated~\cite{N16}. The Langevin equations were solved to find the time-dependent fission paths in the microscopically calculated multidimensional space. It was shown that the peaks positions of the yield depend strongly on the topography of the PES in the pre-scission region, whereas the crucial role in reproduction  of the broadness of the fragment mass distribution play the dissipative collective dynamics and collective inertia.
\section{Summary}

The  following conclusions can be drawn from our investigations:
\begin{itemize}
\item The fission fragment mass distribution of $^{252}$Cf depends weakly on the parity of the initial state. The peaks resulting from propagation of states with negative parity are shifted by 2 mass units to the center of the mass yield in comparison to those obtained from the evolution of positive ones. 
\item There is no strong correlation between fragment mass distribution and energy of propagated eigenstate up to $E_{\rm n}=B_{\rm f}+4$ MeV. The peak position and the broadness of the mass yield is practically independent on the energy of the initial state.
\item The shape of mass yield does not depend on the time of propagation. No difference is observed in the fragment mass distribution at any stage of a time evolution of a particular state. 
\item The mass distributions obtained from the initial states taken as a various combinations of the individual states stay almost unchanged. The resulted peaks are shifted by 4 mass units in comparison to the experimental ones. 
\item This approach is not sufficient to reproduce experimental yields broadness, especially at the high mass asymmetry.  To improve the broadness of the mass yield, the model needs several modifications, e.g. proton and neutron pairing correlations or hexadecapole moment should be taken as the collective coordinates. It would be also worth to check whether the replacement of the perturbative by non-perturbative "cranking" inertia changes the dynamic landscape. It will be a subject of the further investigations.
\end{itemize}
\section*{Acknowledgements}

The Authors are grateful to H. Goutte for the original numerical code. 
This work was supported by the Polish National Science Centre, grant No. 2014/13/N/ST2/02551.

\end{document}